\documentclass[11pt,twoside,twocolumn]{article}

\usepackage{paralist}
\usepackage[affil-it]{authblk}
\usepackage[cm]{fullpage}
\usepackage{amsmath,amssymb,amsfonts}
\usepackage{natbib}
\usepackage{xurl}
\usepackage{graphicx}

\begin{document}

\title{Implementing CUDA Streams into AstroAccelerate -- A Case Study}

\author[1,3]{Jan~Novotn{\'y}}
\author[1,2]{ Karel Ad{\'a}mek}
\author[1]{Wes~Armour}
\affil[1]{Oxford e-Research Centre, Department of Engineering Science, University of Oxford, 7 Keble Rd, Oxford OX1 3QG, United Kingdom.}
\affil[2]{Faculty of Information Technology, Czech Technical University, Th{\'a}kurova 9, 160 00, Prague, Czech Republic.}
\affil[3]{Research Centre for Theoretical Physics and Astrophysics, Institute of Physics, Silesian University in Opava, Bezru{\v c}ovo n{\'a}m. 13, CZ-74601 Opava, Czech Republic.}




  
\maketitle
 
\begin{abstract}
To be able to run tasks asynchronously on NVIDIA GPUs a programmer must explicitly implement asynchronous execution in their code using the syntax of CUDA streams. Streams allow a programmer to launch independent concurrent execution tasks, providing the ability to utilise different functional units on the GPU asynchronously. For example, it is possible to transfer the results from a previous computation performed on input data n-1, over the PCIe bus whilst computing the result for input data n, by placing different tasks in different CUDA streams. The benefit of such an approach is that the time taken for the data transfer between the host (server hosting the GPU) and device (the GPU) can be hidden with computation. This case study deals with the implementation of CUDA streams into AstroAccelerate. AstroAccelerate is a GPU accelerated real-time signal processing pipeline for time-domain radio astronomy.
\end{abstract}

\section{Introduction}
With new coming technological instruments in all fields of science the need to improve computational algorithms, fully utilise hardware architectures, improve softwares, and compete in upcoming data challenges \citep{Gei-Luc:JHPCA:2009:Exascale}, is becoming ever more important. Today, by looking to the worlds most powerful machines \citep{top500,green500}, we can be certain that to reach exaFLOPS performance, heterogeneous computing is necessary. Clear leaders in this area are systems using the power of GPUs, which provide excellent energy efficiency \citep{Mit-etal:ACMCS:2015:GPUsurvey, Ceb-Gue-Gar:IPDPS:2012:efficiency}. To efficiently use such a system we need to ensure that the applications ideally execute functions (kernels) concurrently and the data transfers are hidden by computations. Beginning with CUDA 7,  we can manage this asynchronous behaviour by introducing ``streams''. One of the simplest guides to using this functionality is provided by \citet{Har:2012:blog_Overlap, Har:2015:blog_concurrency}.

In this paper we study the implementation of streams into the AstroAccelerate (AA) project. AA is a GPU-enabled software package that focuses on enabling real-time processing of time-domain radio-astronomy data. It uses the CUDA programming language for NVIDIA GPUs and can perform tasks such as dedispersion, single pulse searching \citep{Adamek_2020} and Fourier Domain Acceleration Searches (FDAS) \citep{Dimoudi_2018,adamek2017improved} in real time on very large data-sets which are comparable to those which will be produced by next generation radio-telescopes such as the Square Kilometre Array (SKA).

\begin{figure}[htbp]
\begin{center}
\includegraphics[width=\linewidth]{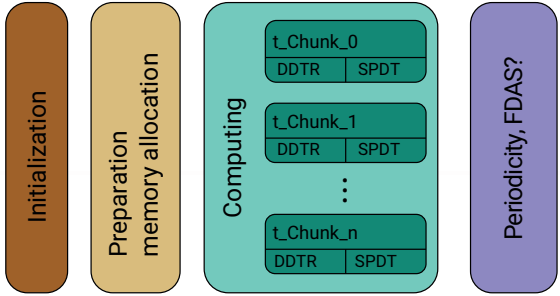}
\caption{\label{fig:AA_schema} Simplified schema of the AstroAccelerate workflow.}
\end{center}
\end{figure}

The AA code can be divided into few main parts as show in Fig.~\ref{fig:AA_schema}. The first part performs the preparation of system and reading user data. The second part consists of mapping tasks to suitable resources and allocation of all necessary memory. The third part, which is the part of the code in which we have implemented streams, is responsible for the dedispersion of data and single pulse searching. The fourth part offers optional features like FDAS (Fourier Domain Acceleration Search) or periodicity searching. 

\begin{figure}[htbp]
\begin{center}
\includegraphics[width=\linewidth]{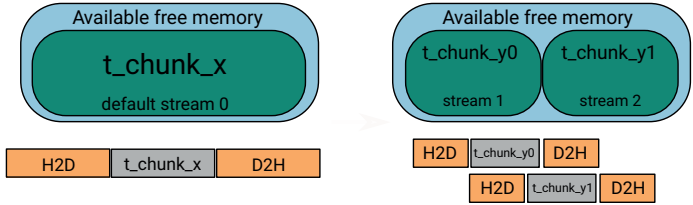}
\caption{\label{fig:AA_stream_ides} Schematic idea of 5the implementation of CUDA streams into AA.}
\end{center}
\end{figure}

To achieve the desired asynchronous behaviour (as shown in Fig.~\ref{fig:AA_stream_ides}, bottom right) of data transfers and computing, we split the input signal to $n$ time chunks (these chunks represent the amount of signal that can fit to GPU memory), and again divide them by the number of desired CUDA streams into smaller chunks. These smaller chunks are then associated with a stream ID. This process is repeated for all time chunks until all data are processed. Care has to be taken to distribute the correct chunk of memory to the correct CUDA stream.

\section{Implementation}
To successfully obtain overlapping data transfers and coherent execution of kernels we perform the following steps:
\begin{inparaenum}[1)]
    \item create CUDA streams;
    \item pinning host memory;
    \item substitute the commands \texttt{cudaMemcpy} to \texttt{cudaMemcpyAsync};
    \item associate streams ID to kernels and memory transfers;
    \item appropriately change all other explicit (wait event commands) and implicit (e.g. memory set, memory allocation) synchronisation commands to non-blocking ones.
\end{inparaenum}

We divide the implementation into two phases: computation of the dedispersion data, and then the single pulse search.\footnote{These parts consists of several kernels, however we name them by their main meaning.} For the correctness of the streams implementation we use visual inspection of the timeline in NVidia Visual Profiler (nvvp). All testing is done on a Tesla V100 GPU.

\subsection{Dedispersion phase}
The host memory is pageable by default, which means that the GPU cannot address the data directly. To be able to overlap kernel execution and data transfers the host memory involved must be pinned -- in CUDA \texttt{cudaMallocHost()} or \texttt{cudaHostAlloc()} is used to allocate pinned memory, to deallocate use \texttt{cudaFreeHost()}.

\begin{figure}[htbp]
\begin{center}
\includegraphics[width=\linewidth]{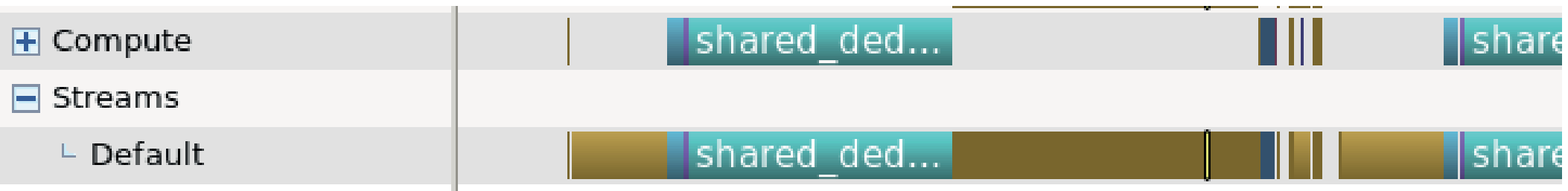}\\[1em]
\includegraphics[width=\linewidth]{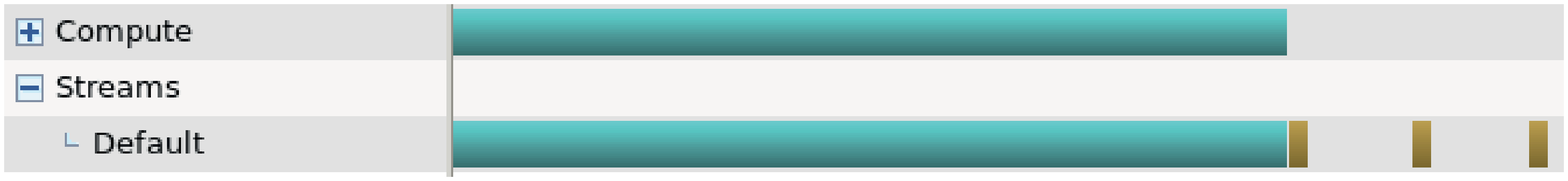}
\caption{\label{fig:nostream} Timeline from nvvp showing the default stream (top) launching data transfers and kernels. A magnified view of the timeline showing the data transfer gaps when pageable memory is used (bottom).}
\end{center}
\end{figure}

When applying the above mentioned points we find that an increase in the throughput of the memory transfers by ${\sim}30\,\%$ can be achieved, along with the benefit of partially overlapping kernels execution.\footnote{The concurrency of kernels can happen when the resources of the GPU are not fully utilised. This can be seen usually at the end of kernel executions when the resources are released.} Another benefit we obtain is the removal of timing gaps between copies from device to host. The gaps are caused by the fact that when the copy is invoked the driver must allocate a temporary page-locked (pinned) host array and transfer the data there (see Fig.~\ref{fig:nostream}). To be precise the time saved is just moved to the allocation and deallocation of memory where we see significant increase (note the increase will be even higher for systems with larger host memory).

\begin{figure}[htbp]
\begin{center}
\includegraphics[width=\linewidth]{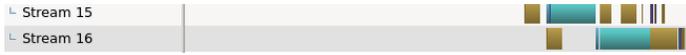}
\caption{\label{fig:stream} Timeline from nvvp showing the CUDA streams working in AA.}
\end{center}
\end{figure}

To decrease time caused by the allocation/deallocation of the host memory we create smaller temporary buffers to move the data from host (big pageable memory) to host (small pinned memory). Using this approach significant time can be saved in the preparation of the host memory. However, the host to host copies block the CUDA streams. A multi-threaded approach can be used to solve this issue, i.e., for every stream ID we create a corresponding CPU thread using OpenMP. The final achieved coherence and overlapping is shown in Fig.~\ref{fig:stream}.

\subsection{Single pulse search phase}
This part of the code deals with single pulse search. In Fig.~\ref{fig:AA_schema} this step is designated as SPDT. As dynamic allocation and deallocation of memory is used in this step which gives rise to synchronous behaviour, we have moved memory allocations from the Computation phase to the Preparation/Memory allocation phase of the code. The final change required was to change or remove the explicit synchronisation events to stream ID synchronisation events with their appropriate stream IDs.

During the implementation of CUDA streams in this phase several problems were encountered. The most crucial is the fact that the \texttt{cudaStreamWaitEvent()} does not work properly when atomic operations are used. This forced us to leave the implementation of CUDA streams in this phase for future work.

\section{Conclusion}
Introducing CUDA streams into sophisticated software like AstroAccelerate was not a straightforward job. We have run into several issues such as stream event barriers not working with atomic operations or significant increases in allocation time for pinned host memory. As a result CUDA streams are used only in the dedispersion phase of our code. We have tested the AA stream version of our code on an SKA-like input signal (1400 MHz central frequency, sampling rate 64 $\mu$s, 4096 channels, 300 MHz bandwidth) and achieved an overall speedup of 1.33 against the non-streams version of AA with a dispersion measure plan computing ${\sim}6000$ trials from 0--3000\,pc\,cm$^{-3}$. 

\section{Acknowledgements}
The authors would like to express their gratitude to the Research Centre for Theoretical Physics and Astrophysics, Institute of Physics, Silesian University in Opava for institutional support, and the support of the OP VVV MEYS funded project CZ.02.1.01/0.0/0.0/16\_019/0000765 "Research Center for Informatics.


\bibliographystyle{unsrtnat}
\bibliography{P8-221}


\end{document}